\documentstyle[aps,multicol,epsf]{revtex}
\begin{document}
\draft
\title{
Modeling Urban Growth Patterns with Correlated Percolation }

\author{Hern\'an A. Makse$^1$, Jos\'e S. Andrade Jr.$^{1,2}$, 
Michael Batty$^3$, \\ Shlomo Havlin$^{1,4}$ and H. Eugene Stanley$^1$}

\address{$^1$Center for Polymer Studies and Dept. of Physics,
 Boston University, Boston, MA 02215 USA}
\address{$^2$Departamento de F\'{\i}sica, Universidade Federal do Cear\'a,
60451-970 Fortaleza, Cear\'a, Brazil}
\address{$^3$Centre for Advanced Spatial Analysis, University College
London, \\ 
1-19 Torrington Place, London WC1E 6BT, UK}
\address{$^4$Department of Physics, Bar-Ilan University,
Ramat-Gan, Israel}

\date{Phys. Rev. E (1 December 1998)}

\maketitle

\begin{abstract}
We propose and test a model that describes the morphology of
cities, the scaling of the urban perimeter of individual cities, and the
area distribution of systems of cities. The model is also consistent
with observable urban growth dynamics, our results agreeing both
qualitatively and quantitatively with urban data. The resulting growth
morphology can be understood from interactions among the constituent
units forming an urban region, and can be modeled using a correlated
percolation model in the presence of a gradient.
\end{abstract}

\begin{multicols}{2}
\narrowtext

\section{Introduction}

Traditional approaches to urban science as exemplified in the work of
Christaller\cite{christaller}, Zipf\cite{zipf}, Stewart and
Warntz\cite{stewart}, Beckmann\cite{beckmann}, 
and Krugman\cite{krugman} are based on the assumption that cities grow
homogeneously in a manner that suggests that their morphology can be
described using conventional Euclidean geometry. However, recent studies
have proposed \cite{fractal} that the complex spatial phenomena
associated with actual urban systems is rather better described using
fractal geometry consistent with growth dynamics in disordered media
\cite{stauffer,percolation1,percolation2}.

Predicting urban growth dynamics also presents a challenge to
theoretical frameworks for cluster dynamics in that different mechanisms
clearly drive urban growth from those which have been embodied in
existing physical models. In this paper, we develop a mathematical model
that relates the physical form of a city and the system within which it
exists, to the locational decisions of its population, thus illustrating
how paradigms from physical and chemical science can help explain a
uniquely different set of natural phenomena - the physical arrangement,
configuration, and size distribution of towns and cities. Specifically,
we argue that the basic ideas of percolation theory when modified to
include the fact that the elements forming clusters are not
statistically independent of one another but are correlated, can give
rise to morphologies that bear both qualitative and quantitative
resemblance to the form of individual cities and systems of cities. Some
of these results are briefly described in Ref.~\cite{hernan}.

We consider the application of statistical physics to urban growth
phenomena to be extremely promising, yielding a variety of valuable
information concerning the way cities grow and change, and more
importantly, the way they might be planned and managed. Such information
includes (but is not limited to) the following:

{\bf \it(i)} the size distributions of towns, in terms of their populations
and areas;

{\bf \it (ii)} the factal dimensions associated with individual cities
and entire systems of cities;

{\bf \it (iii)} interactions or correlations between cities which provide 
insights into their interdependence;

{\bf \it (iv)} the relevance and effectiveness of local planning policies,
particularly those which aim to manage and contain growth.

The size distribution of cities has been a fundamental question in the
theory of urban location since its inception in the late 19th century.
In the introduction to his pioneering book published over 60 years ago,
Christaller \cite{christaller} posed a key question: ``Are there laws
which determine the number, size, and distribution of towns?'' This
question has not been properly answered since the publication of
Christaller's book, notwithstanding the fact that Christaller's theory
of {\it central places\/} \cite{christaller} and its elaboration through
theories such {\it the rank-size rule for cities\/}
\cite{zipf,stewart,beckmann} embody one of the cornerstones of human
geography.

Our approach produces scaling laws that quantify such distributions.
These laws arise naturally from our model, and they are consistent with
the observed morphologies of individual cities and systems of cities
which can be characterized by a number of fractal dimensions and
percolation exponents. In turn, these dimensions are consistent with the
density of location around the core of any city, and thus the theory we
propose succeeds in tying together both intra- and inter-urban location
theories which have developed in parallel over the last 50 years.
Furthermore, the striking fact that cities develop a power law
distribution without the tuning of any external parameter might be
associated with the ability of systems of cities to ``self-organize''
\cite{krugman}.

\section{The DLA Model}

Cities grow in a way that might be expected to resemble the growth of
two-dimensional aggregates of particles, and this has led recent
attempts \cite{fractal,benguigui1,benguigui2} to model urban growth
using ideas from the statistical physics of clusters. In particular,
the model of diffusion limited aggregation (DLA)\cite{dla,vicsek} has
been applied to describe urban growth \cite{fractal}, and results in
tree-like dendritic structures which have a core or ``central business
district'' (CBD).
The DLA model is a physical model used to describe aggregation phenomena
and is related to problems from the field of oil recovery in which
``viscous fingering'' occurs when a low viscosity fluid is pushed under
pressure into a fluid with a larger viscosity (as occurs when an oil
field is flooded with water in an attempt to ``push out the oil'').

The DLA model predicts that there exists only one large fractal cluster
that is almost perfectly screened from incoming ``development units''
(people, capital, resources, etc), so that almost all the cluster growth
occurs in the extreme peripheral tips.
However, quantitative data do not support all the properties of the DLA
model. For instance, the DLA model predicts that the urban population
density $\rho (r)$ decreases from the city center as a power law,
\begin{equation}
\rho(r) \sim
r^{D-2},
\end{equation}
where $r$ is the radial distance from the core, and $D \simeq 1.7$ is
the fractal dimension of DLA. However, urban data have been more
commonly fit to an exponential decay \cite{clark}. In the DLA model
only one large central place or cluster is generated, while a real urban
area is formed by a system of central places that are spatially
distributed in a hierarchy of cities. Still another concern regarding
the morphology of the DLA model is that DLA is a simply-connected
cluster. Cities grow in a more compact way, with a well-defined urban
boundary or external perimeter not accounted for by the dendritic
fractal growth of DLA.

Here we show that an alternative model, in which development units are
correlated rather than being added to the cluster at random, is better
able to reproduce the observed morphology of cities and the area
distribution of sub-clusters (``towns'') in an urban system, and can
also describe urban growth dynamics. Our ``physical'' model
\cite{hernan}, which corresponds to the correlated percolation model
\cite{coniglio,weinrib2,sona,hernan2,hernan3} in the presence of a
density gradient \cite{sapo1,sapo2,sapo3}, is motivated by the fact that
in urban areas development attracts further development. The model
offers the possibility of predicting the global properties (such as
scaling behavior) of urban morphologies.

\section{Correlated Percolation Model}

In the model we now develop, we take into account two points:

{\bf \it (i)} First, data on population density $\rho(r)$ of actual
urban systems are known to conform to the relation \cite{clark}
\begin{equation}
\rho(r) = \rho_0 e^{-\lambda r}, 
\label{rho}
\end{equation}
where $r$ is the radial distance from the central business district
(CBD) situated at the core, and $\lambda$ is the density gradient. The
density gradient quantifies the extent of the urban spread around the
central core. The probability that a unit occupies a given spot
decreases gradually as the distance from a central, compact core
increases.

{\bf \it (ii)} Second, in actual urban systems, the development units
are not positioned at {\it random}. Rather, there exist {\it
correlations} arising from the fact that when a development unit is
located in a given place, the probability of adjacent development units
increases naturally; each site is not independently occupied by a
development unit, but is occupied with a probability that depends on the
occupancy of the neighborhood. In urban settings, development units do
not attach themselves randomly to an existing cluster. Their placement
is strongly influenced by the presence of other units. When a unit
occupies a certain location, the probability of additional development
is highest in its vicinity, and this probability decreases at a certain
rate as the distance from the unit increases. Thus, the rules of
placement are affected by long-range ``interactions'' that influence how
clusters form and grow. What happens at a given site depends on the
state of many other sites. These correlations reflect the tendency of
people to locate next to one another, as articulated in traditional
urban science as economies of urban agglomeration.

In order to quantify these ideas, we consider the {\it correlated}
percolation model \cite{coniglio,weinrib2,sona,hernan2,hernan3} in the
presence of a concentration gradient \cite{sapo1,sapo2,sapo3}. We start
by describing the uncorrelated site percolation problem, which
corresponds to the limit where correlations are so small as to be
negligible \cite{stauffer,percolation1,percolation2}. We first define a
random number $u$({\bf\it r}), called the occupancy variable, at every
site {\bf\it r}=$(i,j)$ in a square lattice of $L^2$ sites. The numbers
$u$({\bf\it r}) are uncorrelated numbers  with a uniform
probability distribution between $[0,1]$. A site in the lattice is
occupied if the occupancy variable $u$({\bf\it r}) is smaller than the
occupation probability $p$, which is a quantity fixed for every site in
the lattice. A cluster is a set of sites connected via nearest neighbor
sites. When $p$ is small only isolated clusters exist. At a critical
value of the concentration called $p_c$ an ``incipient infinite
cluster'' forms which, for a finite system, connects two sides of the
system.

Our basic model is a percolation model modified to introduce
correlations among the units and the fact that the concentration $p$ is
not constant for all the points in the lattice but presents the gradient
shown in Eq. (\ref{rho}). In our model we consider ``development
units'' representing buildings, people, resources which are added to the
cluster in similar fashion as in percolation. Since these units are
added in a correlated fashion, we next consider a modification of the
percolation problem to incorporate correlations among the occupancy
variables $u$({\bf\it r}).

To introduce correlations among the variables we use a method proposed
in \cite{hernan3} which is a modification of the Fourier filtering
method (Ffm) \cite{ffm1,ffm2,ffm3,sona} suitable for large systems. Consider a
stationary sequence of $L^2$ uncorrelated random numbers $\{ u(r)\}$,
$r=(i,j), i,j=1,..,L$. The correlation function is $ \langle u(r) ~
u(r') \rangle \sim \delta_{r,r'}$, with $\delta_{r,r'}$ the Kronecker
delta, and the brackets denote an average with respect to a Gaussian
distribution. We use the sequence $ \{ u(r) \} $, to generate a new
sequence, $ \{ \eta(r) \}$, with a long-range power-law correlation
function $C(\ell)$ of the form \cite{hernan3}
\begin{equation}
C(\ell) \equiv  \langle u(r) ~
u(r') \rangle = (1+\ell^2)^{-\alpha/2},
\label{coreal}
\end{equation}
where, $\ell = |r-r'|$, $\alpha$ is the correlation exponent, and the
long-range correlations are relevant for $0<\alpha<d=2$, where $d$ is
the dimension of the substrate--- $\alpha\ge 2$ corresponds to the 
uncorrelated problem, and $\alpha\to 0 $ to the strongly correlated problem.

The spectral density $S(q)$, defined as
the Fourier transform of $C(\ell)$, has the form
\begin{equation}
\label{analfourier2d}
S(q) = \frac{2\pi}{\Gamma(\beta_2+1)} ~ \left( \frac{q}{2}
\right)^{\beta_2} ~ K_{\beta_2}(q),
\end{equation}
where $q=|\vec{q}|$, $q_i=2\pi m_i/L$, $-L/2 \leq m_i \leq L/2$,
$i=1,2$, and $\beta_2=(\alpha-2)/2$. $ \{ \eta(q) \}$ are 
the Fourier transform coefficients of $\{ \eta(r) \}$, and satisfy
\begin{equation}
\label{fourierg}
\eta(q) = \left(S(q)\right)^{1/2} ~ u(q) ,
\end{equation}
where $\{u(q)\}$ are the Fourier transform coefficients of $\{u(r)\}$.

The actual numerical algorithm for Ffm consists of the following steps:
{\bf \it (i)} Generate a two-dimensional sequence $ \{ u(r) \} $ of
uncorrelated random numbers with a Gaussian distribution, and calculate
the Fourier transform coefficients $\{u(q)\}$. {\bf \it (ii)} Obtain
$\{\eta(q)\}$ using (\ref{analfourier2d}) and (\ref{fourierg}). {\bf \it
(iii)} Calculate the inverse Fourier transform of $\{ \eta(q) \}$ to
obtain $\{\eta(r)\}$, the sequence in real space with the desired
power-law correlation function which asymptotically behaves as
\begin{equation}
\label{assim}
C(\ell) \sim \ell^{-\alpha}.
\end{equation}

The assumption of power-law interactions is motivated by the fact that
the ``decision'' for a development unit to be placed in a given location
decays gradually with the distance from an occupied neighborhood.

Finally we consider that the development units are positioned with a
probability which behaves in the same fashion as known for cities Eq. 
(\ref{rho}).
Therefore we relax the assumption that the concentration $p$ is constant
for all the points in the lattice, and we consider that the development
units are positioned with an occupancy probability
\begin{equation}
p(r) \equiv \rho(r)/\rho_0,
\label{gradient}
\end{equation}
that behaves in the same fashion as is known in observations of real
cities. This last modification corresponds to the percolation problem in
the presence of a concentration gradient proposed in
\cite{sapo1,sapo2,sapo3}.

\vspace{-3.5cm}
\begin{figure}
\centerline{ 
\vbox{ \hbox{\epsfxsize=12.cm \epsfbox{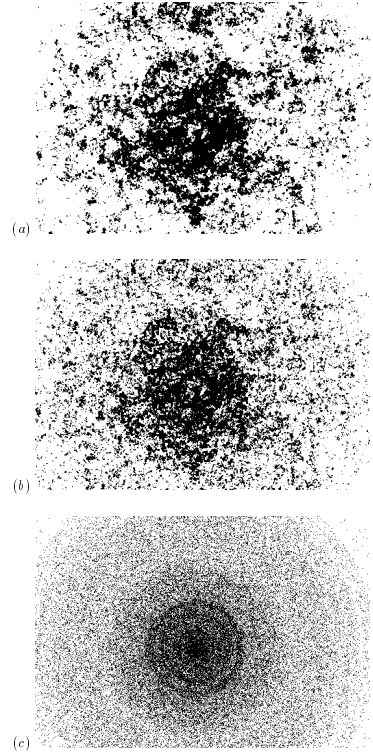}
    }
  }
}
\vspace{-2.5cm}
\caption{Simulations of urban systems 
for different degrees of correlation. Here, the urban areas are red,
and the external perimeter or urban boundary of the largest cluster
connected to the CBD is light green. In all the figures, we fix the
value of the density gradient to be $\lambda=0.009$. $(a)$ and $(b)$
Two different examples of interactive systems of cities for correlation
exponents $\alpha=0.6$ and $ \alpha=1.4$, respectively. The development
units are positioned with a probability that decays exponentially with
the distance from the core. The units are located not randomly as in
percolation, but rather in a correlated fashion depending of the
neighboring occupied areas. The correlations are parametrized by the
exponent $\alpha$. The strongly correlated case corresponds to small
$\alpha$ ($\alpha \to 0$). When $\alpha>d$, where $d$ is the spatial
dimension of the substrate lattice ($d=2$ in our case), we recover the
uncorrelated case. Notice the tendency to more compact clusters as we
increase the degree of correlations ($\alpha \to 0$). $(c)$ As a zeroth
order approximation, one might imagine the morphology predicted in the
extreme limit whereby development units are positioned at {\it random},
rather than in the correlated way of Figs. $1a$ and $1b$. The results
for this crude approximation of a non-interactive (uncorrelated) system
of cities clearly display a drastically different morphology than found
from data on real cities (such as shown in Fig. \protect\ref{dy}$a$). The
non-interactive limit looks unrealistic in comparison with real cities,
for the lack of interactions creates a urban area characterized by many
small towns spread loosely around the core.}
\label{static}
\end{figure}

In order to apply the above procedure to the percolation problem, we study
the probability distribution $P(\eta)$ of the correlated sequence
$\eta(r) $. We find that when the uncorrelated variables $u(r)$ are
taken from a Gaussian distribution, $\eta(r)$ also has a Gaussian
distribution. We next discretize the variables generating a sequence
$\mu(i,j)$, according to $\mu(i,j) =
\Theta ( \theta - \eta(r) ) $ where $\theta$ is chosen to satisfy
$ p(r) = \int_{-\infty}^{\theta} P(\eta) d\eta $, with $p(r)$ the occupancy
probability and $\Theta$ is the Heaviside step function.

Notice that we have defined two different properties of the system.
First we introduced long range correlations among the variables by
modifying the occupancy variables $\eta(r)$. These correlations are
isotropic, i.e., all the points in space are connected by interactions
quantified by a power law. The fact that we consider a slowly-decaying
power-law scale-free function is due to the fact that any other
correlation function will display a cut-off after which correlations are
negligible. Since we are looking at properties of actual cities at large
length scales, a coarse grain will transform a finite range correlated
system into a uncorrelated system, i.e., a system with a finite cut-off
in the correlations becomes uncorrelated at large scales. This situation
occurs when we consider power law correlations of the form
(\ref{assim}) since it is a scale-free function. Thus correlations are
expected to be relevant at all length scales. One must distinguish the
type of correlation introduced by (\ref{assim}) from the correlations
arising at the critical concentration $p_c$. In this case, the
connectedness length of the system is said to be infinite since two
occupied sites separated by an arbitrary distance may be connected by
the infinite cluster, and thus correlated in space. However, the
correlations introduced by (\ref{assim}) goes beyond this type of
connection between sites. Due to correlation of type (\ref{assim}), even
sites which belong to different clusters are correlated.

Second, we consider that the probability of occupancy of the sites
decays exponentially with the center point always occupied. This property
of the system is independent of the type of correlation chosen. The
correlation exponent $\alpha$ and the density gradient $\lambda$ are the
only parameters of the model to be determined by empirical observations.

\section{Statics}

We first discuss the influence of the correlations on the morphology of
a system of cities generated in the present model. Therefore we fix the
value of the concentration gradient $\lambda$ in Eq. (\ref{gradient})
and we show in Fig. \ref{static} our simulations of correlated urban
systems for different degrees of correlation. We see that the larger
the degree of correlations the more compact the clusters are. The
correlations have the effect of agglomerating the units around an urban
area. In the simulated systems the largest city is situated in the core
(which acts as the ``attractive'' center of the city), and this is
surrounded by small clusters or ``towns.'' The correlated clusters are
fairly compact near their respective centers and become less compact
near their boundaries, in qualitative agreement with empirical data on
actual large cities such as Berlin, Paris and London. (see i.e. Refs.
\cite{fractal,berlin}). The strongly correlated case of Fig. \ref{static}a
$\alpha\to0$ results in a system of cities looking more 
realistic than the uncorrelated case (Fig. \ref{static}c). The uncorrelated 
case results in a systems of very small cities spread around a central city,
while the cities in the correlated case look more compact and more realistic.

\vspace{-2.5cm}
\begin{figure}
\centerline{
\hbox{ 
\epsfxsize=12.0cm \epsfbox{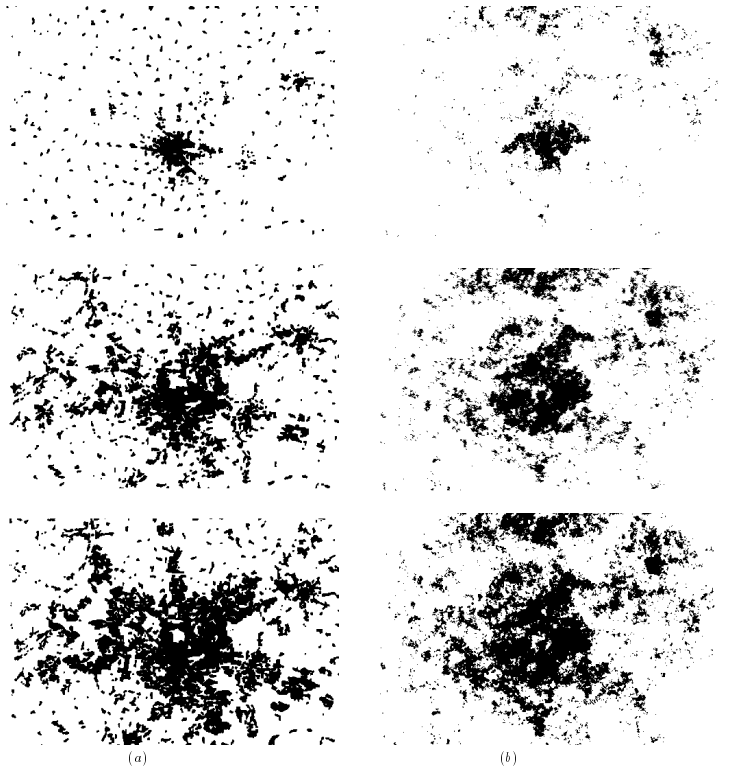} 
}
}
\vspace{-2.5cm}
\caption{Qualitative comparison between the actual urban data and the
proposed model. $(a)$ Three steps of the growth with time of Berlin and
surrounding towns. Data are shown for the years 1875, 1920, and 1945
(from top to bottom). $(b)$ Dynamical urban simulations of the proposed
model. We fix the value of the correlation exponent to be $\alpha =
0.05$ (strongly correlated case), and choose the occupancy probability
$p(r)$ to correspond to the density profiles shown in Fig.
\protect\ref{berdensity}. We use
the same seed for the random number generator in all figures.}
\label{dy}
\end{figure}

For distances smaller that $r_f$, there is a high concentration of sites
since $p(r)>p_c$, and the cluster thus generated plays the role of the
infinite cluster. For distance larger than $r_f$, we have $p(r)<p_c$,
so that only isolated clusters exist, which form the system of small
cities surrounding the large city situated in the core. 

The urban boundary of the largest city is defined to be the external
perimeter of the cluster connected to the CBD. Since $p(r)$ decreases
as one moves away from the core, the probability that the largest
cluster remains connected decreases with $r$. The {\it mean} distance
of the perimeter from the center $r_f$ is determined by the value of $r$
for which $p(r)$ equals the percolation threshold---i.e. $p(r_f)=p_c$,
so \cite{sapo1,sapo2,sapo3}
\begin{equation}
r_f = \lambda^{-1}\ln(1/p_c).
\end{equation}

The geometrical properties of the external perimeter of the largest city
correspond to the properties of the external perimeter of the infinite
cluster of the percolation problem in the absence of a gradient
\cite{sapo1}. The critical properties of the clusters can be analyzed
in terms of the percolation properties. Percolation clusters formed
below $p_c$ are characterized by a finite connectedness length which is
the typical distance at which two sites are expected to be connected via
nearest neighbor sites (do not confuse with the correlations introduced
via Eq. (\ref{assim})). This connectedness length diverges when the
infinite cluster forms at $p_c$, i.e. $\xi \sim |p-p_c|^{-\nu}$, where
$\nu$ is the connectedness length exponent. In the case of gradient
percolation the clusters formed below $p_c$ for $r>r_f$ are
characterized by this length, which is now a function of the distance
$r$
\begin{equation}
\xi(r) \sim |p(r)-p_c|^{-\nu}.
\label{conn}
\end{equation}
Moreover, due to the existence of long range correlation among the
variables the exponent $\nu$ is not universal, but changes continuously
with the degree of correlation given by $\alpha$ \cite{sona}. We will
see that several critical properties of the percolation clusters change
with the correlations.

The width
$\sigma_f$ of the external perimeter of the largest city is defined as
\begin{equation}
\sigma_f \equiv\langle(r-r_f)^2\rangle^{1/2},
\label{m2}
\end{equation}
where $r_f\equiv\langle r \rangle$, 
and $r$ belongs to the external perimeter of the
central cluster.
The width of the external perimeter is a function of the concentration
gradient $\lambda$ and it is known to scale as \cite{sapo1}
\begin{equation}
\sigma_f \sim \lambda^{-\nu/(1+\nu)}
\label{sigma}
\end{equation}

The value of $\nu$ corresponding to the uncorrelated percolation problem
is $\nu=4/3$. However the presence of long range correlations of the
type (\ref{assim}) drastically affects the value of the connectedness
exponent, which is now a function of $\alpha, \nu(\alpha)$ as observed
in previous studies of long range correlated percolation
\cite{weinrib2,sona}. We have simulated the correlated percolation
problem with a gradient and using Eq. (\ref{sigma}) we find a drastic
increase of $\nu(\alpha)$ with the increase of the long-range
correlations $(\alpha\to 0)$ (Figs.~\ref{nub}a, \ref{nub}b). In particular
$\nu(\alpha)$ seems to increase very drastically for a system of strong
correlations $\alpha\to0$. In fact for such a system, we expect a mean
field situation where all sites in the lattice are connected to the rest of
the sites. In this case the percolation threshold for the site
percolation problem should be $p_c=0.5$ and the connectedness length
should be zero below $p_c$ and infinite above $p_c$.

The scaling of the length of the urban boundary of the largest city
within a region
of size $\ell$,
\begin{equation}
L(\ell)\sim\ell^{D_e},
\end{equation}
defines the fractal dimension $D_e$, which we calculate to have values
$D_e \simeq 1.33$ for the uncorrelated case, and $D_e\simeq1.4$ for
strong correlations $(\alpha\to 0)$ (Fig. \ref{nub}c).
 The small variation of the fractal
dimension of the external perimeter does not rule out the fact that it
may be independent of the correlations. These values are consistent
with actual urban data, for which values of $D_e$ between $1.2$ and
$1.4$ are measured \cite{fractal}. 

Near the frontier and on length scales smaller than the width of the
frontier $\sigma_f$, 
the largest cluster has fractal dimension $d_f\simeq 1.89$, as
defined by the ``mass-radius'' relation
\begin{equation}
M(r)\sim r^{d_f},
\end{equation}
where $M(r)$ is the mass of the cluster inside a region of radius $r$. 
The value $d_f \simeq 1.89$ corresponds to the fractal
dimension of the uncorrelated percolation clusters and we find that it
is valid independent of the correlations. However, as $\alpha\to 0$ we
expect a compact cluster with dimension $d_f=2$. The fact that we are
unable to see this limit might be due to numerical limitations 
near the
mean field point $\alpha=0$.

The number of sites of the frontier $N_f$ also scales with the
concentration gradient \cite{sapo1}
\begin{equation}
N_f \sim \lambda^{-\nu(d_f-1)/(1+\nu)}.
\end{equation}
This relation provides another way of calculating the fractal dimension
$d_f$ and the exponent $\nu$, which we used to verify our
calculations.

It is important to stress that under the present percolation picture
cities are fractal structures only near the external perimeter of the
largest city, and on length scales smaller than the width of the
frontier defined by Eq. (\ref{m2}). The width is a function of the
concentration gradient $\lambda$, Eq. (\ref{sigma}) 
so that the larger the spread of the
city the larger the region where the city is fractal. However, at
distances close to the center of the largest city, the cluster is
clearly non fractal since $p(r) > p_c$ and the cluster becomes compact.
On the other side for larger distances $p(r)<p_c$, only small isolated
clusters exist with a definite connectedness length associated with them
(\ref{conn}), so that they are not fractal either.

\section{Area distribution of urban settlements}

So far, we have argued how correlations between occupancy probabilities
can account for the irregular morphology of towns in a urban system. As
can be seen in Fig. \ref{dy}$a$, the towns surrounding a large city like
Berlin are characterized by a wide range of sizes. We are interested in
the laws that quantify the town size distribution $N(A)$, where $A$ is
the area occupied by a given town or ``mass'' of the agglomeration.

\begin{figure}
\centerline{ \vbox{ \hbox{\epsfxsize=7.0cm \epsfbox{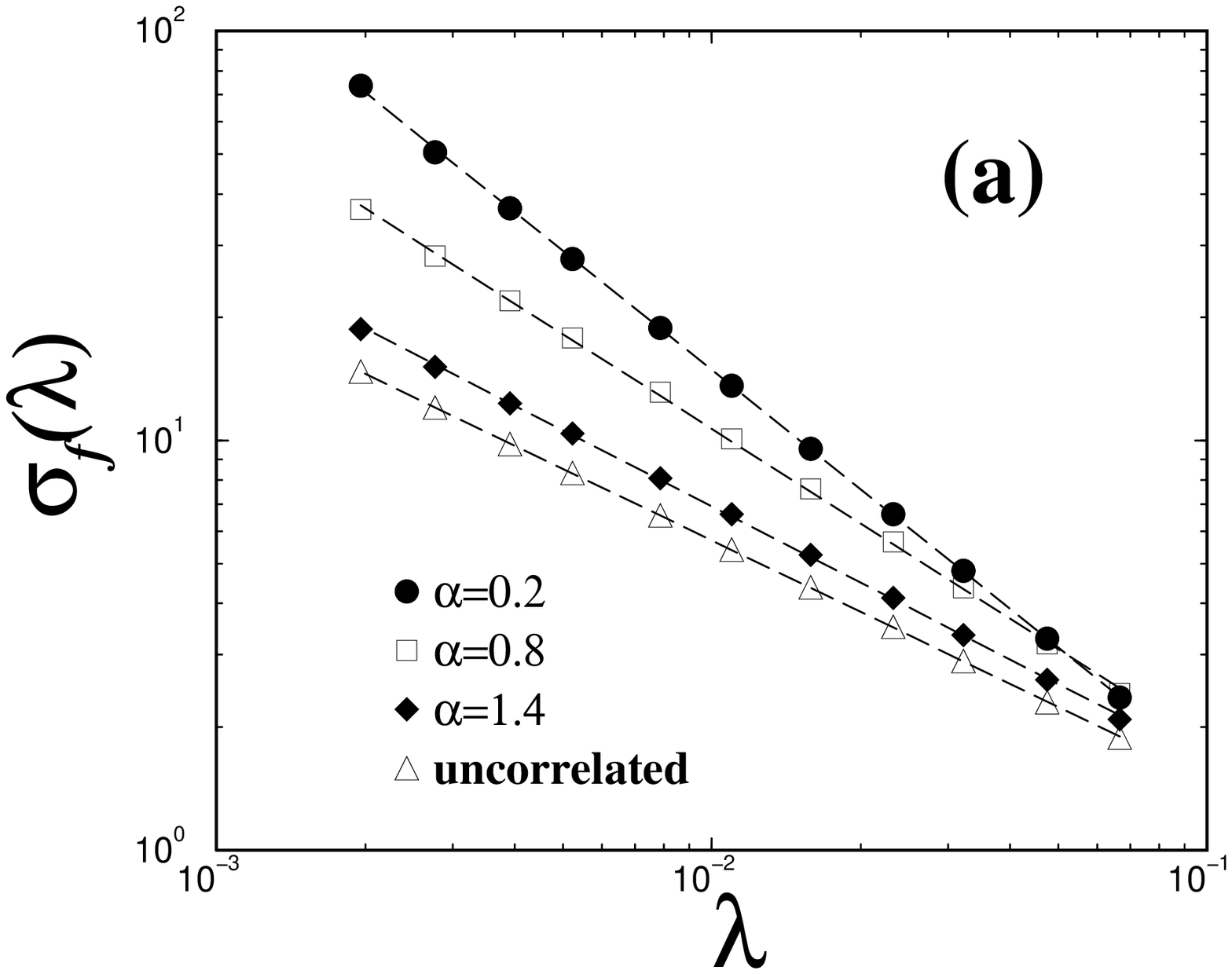} }}}
\centerline{ \vbox{ \hbox{\epsfxsize=7.0cm \epsfbox{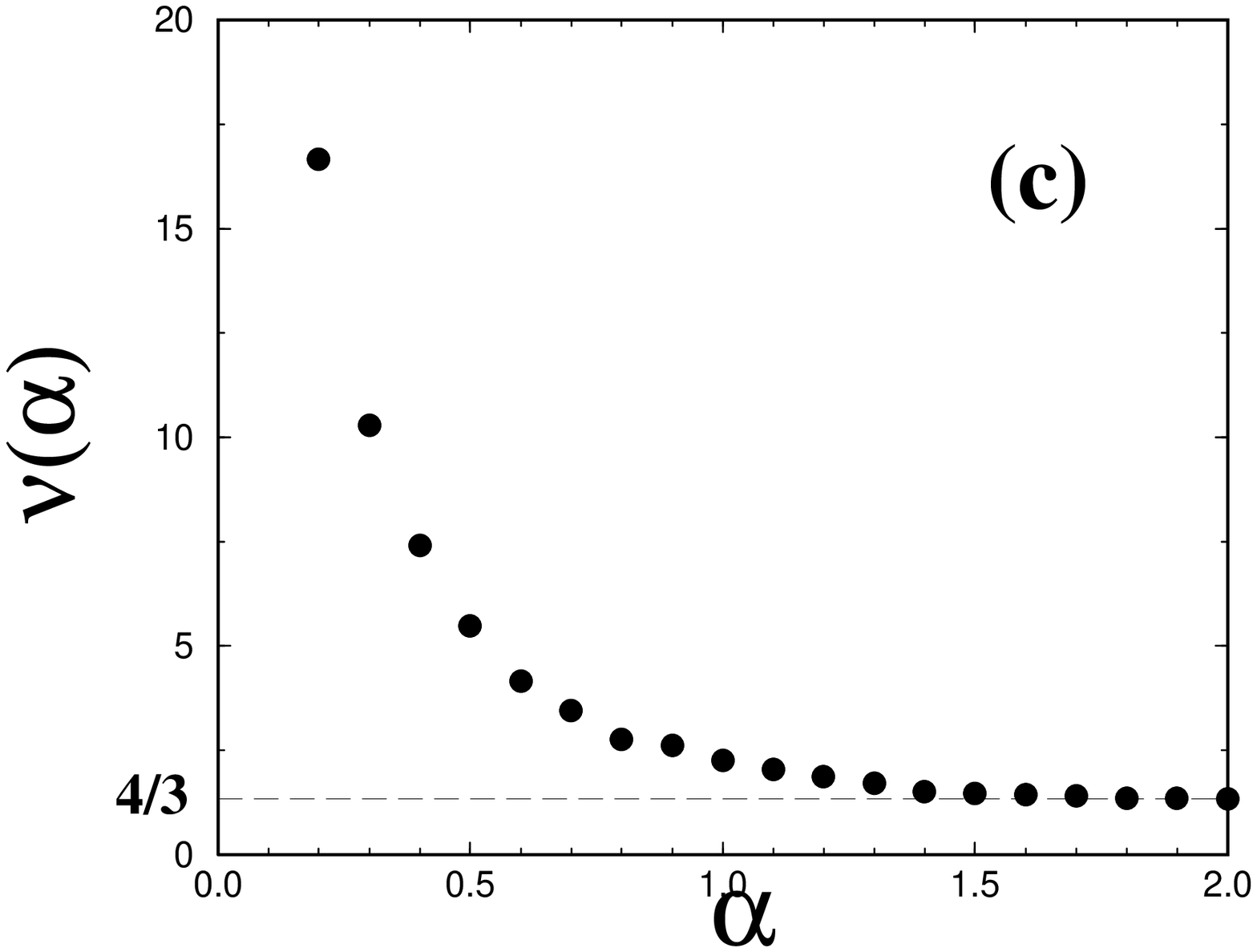} }}}
\centerline{ \vbox{ \hbox{\epsfxsize=7.0cm \epsfbox{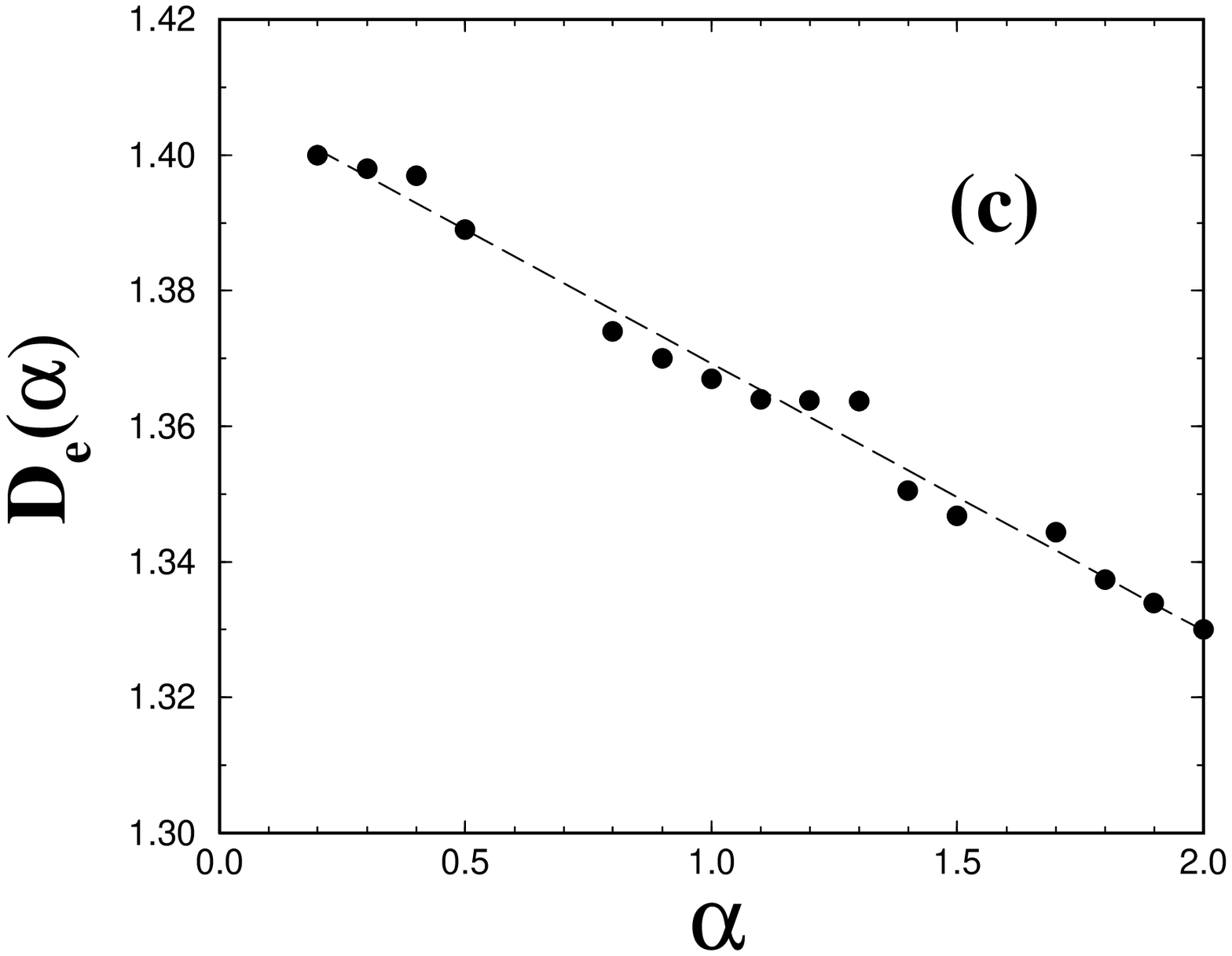} }}}
\caption{
(a) Width of the external perimeter as a function of the density 
gradient, $\sigma_f(\lambda)$, for several degree of correlations. 
(b) Connectedness length exponent $\nu(\alpha)$ as a function of the
correlation exponent $\alpha$ calculated from Fig. \protect\ref{nub}a
using Eq. (\protect\ref{sigma}). The value $\nu=4/3$ corresponds
to the uncorrelated percolation problem ($\alpha=2$).
(c) Fractal dimension of the external perimeter of the largest 
cluster as a function of the degree of correlation, $D_e(\alpha)$.
}
\label{nub}
\end{figure}

We have analyzed the distribution of areas of actual cities, such as the
system of cities surrounding London and Berlin for different years 
(Fig. \ref{dy}a), and
also we analyzed the area distribution of urban systems at larger scales
by using the data of all settlements of Great Britain for the years
1981 and 1991 
 \cite{web}. 
In the case of the towns around
Berlin and London, we first digitize the empirical data of Fig. 4.1 of
Ref.~\cite{berlin} (Berlin 1920 and 1945, shown in the last two panels
of Fig. \protect\ref{dy}$a$), and Fig. 10.8 of Ref.
\protect\cite{fractal} corresponding to London 1981. Then, we count the
number of towns that are covered by $A$ sites, putting the result in
logarithmically spaced bins (of size $1.2^k$, with $k=1, 2,...$), and
averaging over the size of the bin. 

We calculate the actual distribution of the areas of the urban
settlements around Berlin and London, and find (Fig.~\ref{nua}a) that
for both cities, $N(A)$ follows a power-law with exponent close to $2$
\begin{equation}
N(A) \sim A^{-1.98}\qquad\mbox{[Berlin, 1920, 1945]},
\end{equation}
\begin{equation}
N(A) \sim A^{-1.96}\qquad\mbox{[London, 1981]}.
\end{equation}
Figure \ref{nua}b shows the distribution of all urban areas in Great Britain
for the years 1981 and 1991..
We find a power-law with an exponent consistent with the data of London
and Berlin at smaller scales: 
\begin{equation}
N(A) \sim A^{-2.03}\qquad\mbox{[Great Britain, 1981, 1991]}.
\end{equation}
Other studies have recently confirmed the
validity of these results for larger length scales and also for the population
distributions which is known to scale as
the occupied area \cite{zanette}.

These results can be understood in the context of our model. Insight into
this distribution can be developed by first noting that the small
clusters surrounding the largest cluster are all situated at distances
$r$ from the CBD such that $p(r) < p_c$ or $r > r_f$. Therefore, we find
$N(A)$, the cumulative area distribution of clusters of area $A$, to be
\begin{equation}
\label{distribution}
N(A) = \int_0^{p_c} n(A,p) ~ dp \sim A^{-(\tau+1/d_f \nu)}.
\end{equation}
Here, 
\begin{equation}
n(A,p) \sim A^{-\tau}g(A/A_0)
\end{equation}
is defined to be the average number of clusters containing $A$ sites for
a given $p$  at a fixed distance $r$, and $\tau=1+2/d_f$. Here,
\begin{equation}
A_0(r) \sim \xi(r)^{d_f} \sim |p(r)-p_c|^{-d_f\nu}
\end{equation}
corresponds to the maximum typical area occupied by a cluster situated
at a distance $r$ from the CBD, while $g(A/A_0)$ is a scaling function
that decays rapidly (exponentially) for $A>A_0$.

\begin{figure}
\centerline{ \vbox{ \hbox{\epsfxsize=8.0cm \epsfbox{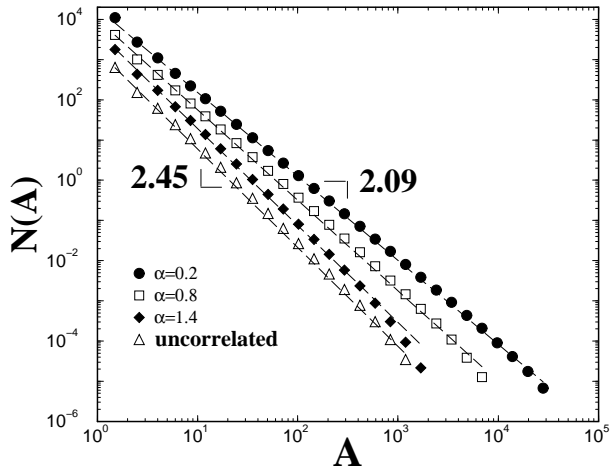} }}}
\caption{ 
Log-log plot of the area distribution function $N(A)$ calculated for the
present model for different degrees of correlation. From top to bottom,
$\alpha=0.2$, $\alpha=0.8$, $\alpha=1.4$, and uncorrelated case. The
linear fits correspond to the predictions of
Eq.~(\protect\ref{distribution}) using the values of $\nu(\alpha)$ from
Fig.~\protect\ref{nub}b, and $d_f=1.89$.}
\label{nuc}
\end{figure}

\begin{figure}
\centerline{ \vbox{ \hbox{\epsfxsize=7.0cm \epsfbox{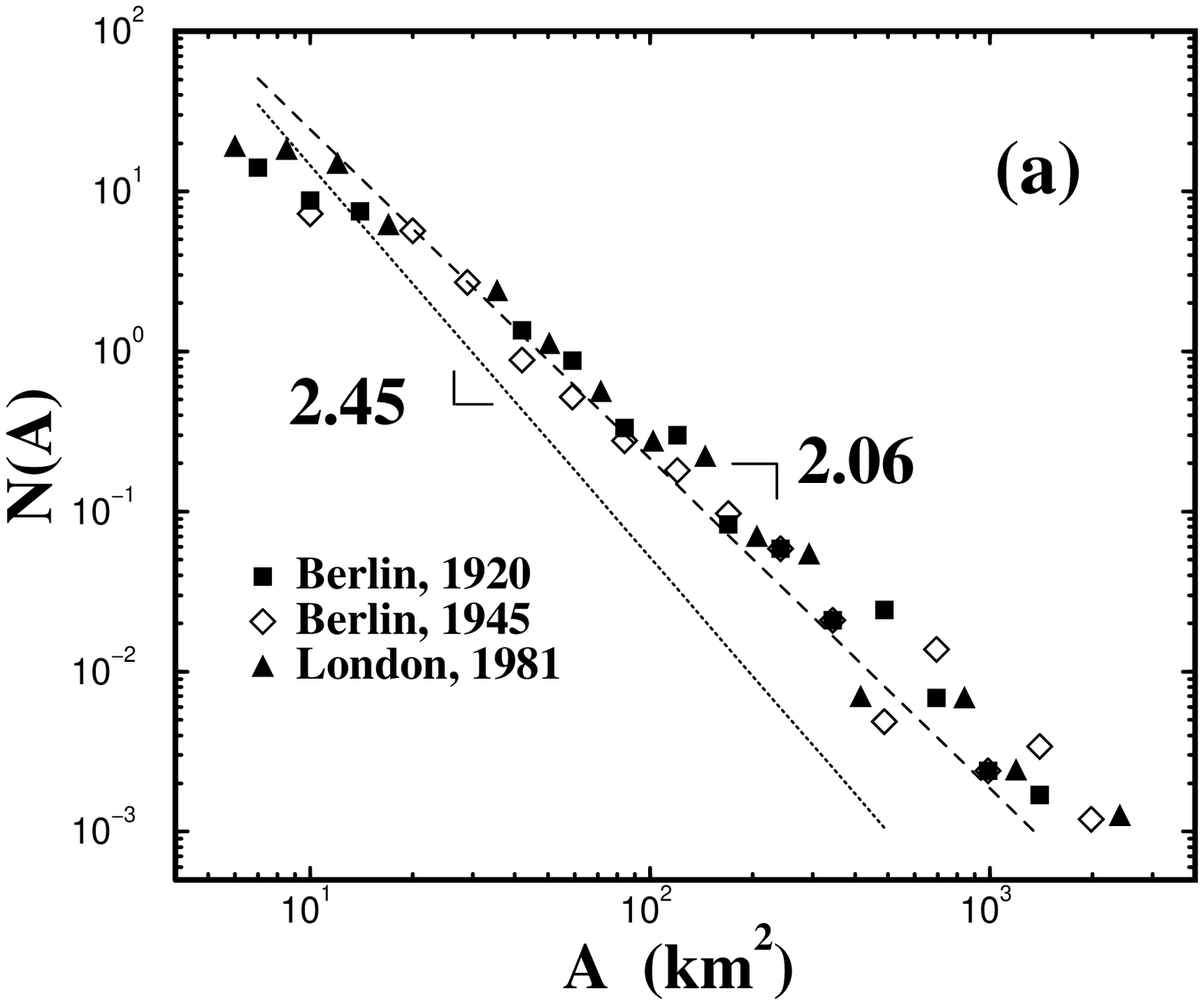} }}}
\centerline{ \vbox{ \hbox{\epsfxsize=7.0cm \epsfbox{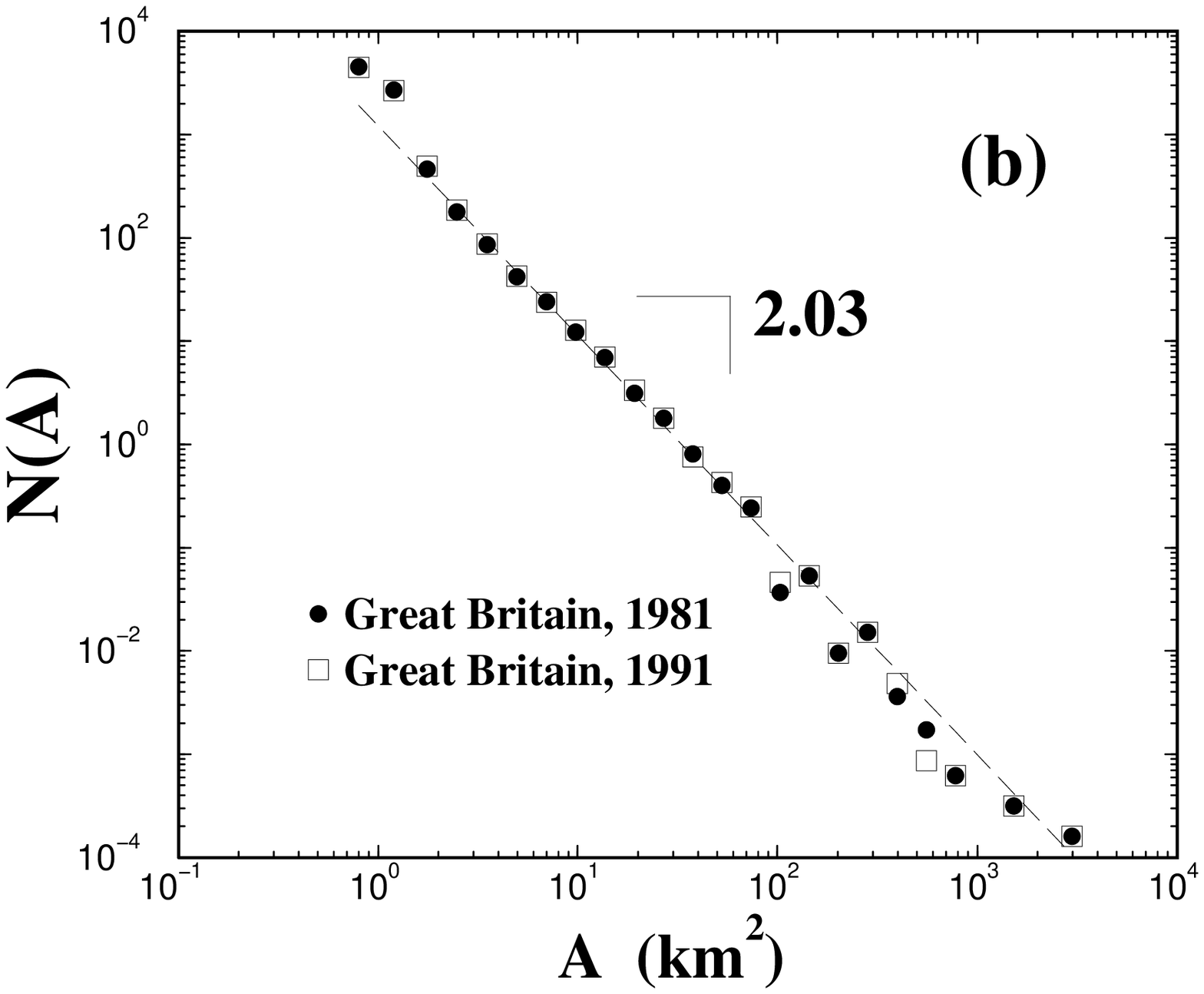}
 }}}
\vspace{1cm}
\caption{(a) 
Log-log plot of the area distribution $N(A)$ of the actual towns around
Berlin and London. We first digitize the empirical data of Fig. 4.1 of
Ref.~\protect\cite{berlin} (Berlin 1920 and 1945, shown in the last two
panels of Fig. \protect\ref{dy}$a$), and Fig. 10.8 of
Ref.~\protect\cite{fractal} (London 1981). Then, we count the number of
towns that are covered by $A$ sites, putting the result in
logarithmically spaced bins (of size $1.2^k$, with $k=1, 2,...$), and
averaging over the size of the bin. A power-law is observed for the area
distributions of both urban systems. The dotted line shows the
predictions of our model for the uncorrelated case (slope$=2.45$), while
the dashed line gives results for the strongly correlated case
(slope$=2.06$). Note that the area distributions for both cities agree
much better with the strongly correlated case ($\alpha\to 0$). (b)
Log-log plot of the area distribution of all the urban areas in 
Great Britain in 1981 and 1991. The data fits to a power law of exponent
2.03. Notice also the very small changes of the urban areas in a 10
year period.}
\label{nua}
\end{figure}

In our numerical simulations we find a drastic increase of $\nu(\alpha)$
with the increase of the long-range correlations $(\alpha\to 0)$
(Fig.~\ref{nub}b) The connectedness exponent $\nu(\alpha)$ affects
the area distribution of the small clusters around the CBD (Fig.
\ref{nuc}), as specified by Eq. (\ref{distribution}), and can be used
to quantify the degree of interaction between the CBD and the small
surrounding towns. For instance, for a strongly correlated system of
cities characterized by small $\alpha$, $\nu(\alpha)$ is large so that
the area $A_0(r)$ and the linear extension $\xi(r)$ of the towns will be
large even for towns situated away from the CBD. This effect is observed
in the correlated systems of cities of Fig.
\ref{static}.


In Fig. \ref{nua}a we plot the power-law for the area distribution
predicted by Eq. (\ref{distribution}) along with the real data for
Berlin and London and all Great Britain. In particular, the slope predicted
for the uncorrelated system is
\begin{equation}
N(A) \sim A^{-2.45}, \qquad\mbox{[uncorrelated model]}, 
\end{equation} 
while for the strongly correlated model is
\begin{equation}
N(A) \sim A^{-2.06}, \qquad\mbox{[strongly correlated model]}.
\end{equation} 

Therefore, we find that the power law of the 
area  distribution of actual cities are
consistent with the prediction (dashed line, Fig. \ref{nua}a) 
for the case of highly
correlated systems. These results quantify the qualitative agreement
between the morphology of actual urban areas and the strongly correlated
urban systems obtained in our simulations. Clearly, the exponent of the
area distribution provides a stronger test of our model against
observations than does the fractal dimension $D_e$ of the perimeter.

\section{Dynamics}

\begin{figure}
\centerline{ \vbox{ \hbox{\epsfxsize=7.0cm \epsfbox{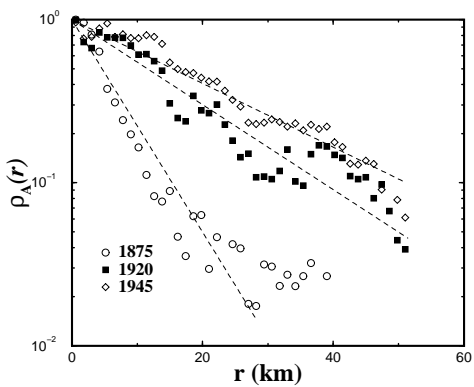} }}}
\caption{Semi-log plot of the 
density of occupied urban areas $\rho_A(r)=e^{-\lambda r}$ for the three
different stages in the growth of Berlin shown in Fig. 
\protect\ref{dy}$a$. Least
square fits yield the results $\lambda \simeq 0.030$, $\lambda \simeq
0.012$, and $\lambda \simeq 0.009$, respectively, showing the decrease
of $\lambda$ with time. We use these density profiles in the dynamical
simulations of Fig. \protect\ref{dy}$b$.}
\label{berdensity}
\end{figure}

We now discuss a generalization of our static model to describe the
dynamics of urban growth. Empirical studies \cite{clark} of the
population density profile of cities show a remarkable pattern of
decentralization, which is quantified by the decrease of $\lambda(t)$
with time (see Table 4 in Ref. \cite{mills}, and Fig. \ref{berdensity}). 
Therefore the
 dynamics in the model
are quantified by a decreasing $\lambda(t)$, as occurs in actual urban
areas. In the context of our model, this flattening pattern can be
explained as follows. The model of percolation in a gradient can be
related to a dynamical model of units (analogous to the development
units in actual cities) diffusing from a central seed or core
\cite{sapo1,sapo2,sapo3}. In this dynamical system, the units are
allowed to diffuse on a two-dimensional lattice by hopping to
nearest-neighbor positions. The density of units at the core remains
constant: whenever a unit diffuses away from the core, it is replaced by
a new unit. 

The density of units can be mapped to the density of
occupied urban areas
\begin{equation}
\rho_A(r)=e^{-\lambda r},
\end{equation}
which in turn is proportional to the {\it population} density $\rho(r)$
\cite{fractal}. A well-defined diffusion front, defined as the boundary
of the cluster of units that is linked to the central core, evolves in
time. The diffusion front corresponds to the urban boundary of the
central city. The static properties of the diffusion front of this
system were found to be the same as those predicted by the gradient
percolation model
\cite{sapo1,sapo2,sapo3}. Moreover, the dynamical model can explain 
the decrease of $\lambda(t)$ with time observed empirically. As the
diffusion front situated around $r_f$ moves away from the core, the city
grows and the density gradient decreases since $\lambda(t) \propto
1/r_f$.

These considerations are tested in Fig.~\ref{dy}$b$, which shows our
dynamical urban simulations of a strongly interacting system of cities
characterized by a correlation exponent $\alpha = 0.05$ for three
different values of $\lambda$ obtained from the data of Berlin from
Fig. \ref{berdensity}. 
Qualitative agreement is observed
between the morphology of the cities and towns of the actual data of
Fig. \ref{dy}$a$ and the simulations of Fig. \ref{dy}$b$.

\section{Discussion: Urban planning}

Throughout this century, the dominant planning policy in many western
nations has been the containment of urban growth. This has been effected
using several instruments, particularly through the siting of new
settlements or new towns at locations around the growing city which are
considered to be beyond commuting distance, but also through the
imposition of local controls on urban growth, often coordinated
regionally as ``green belts'' \cite{munton}. One of the key elements in
the growth models we have proposed here is the characteristic length
scale over which growth takes place. In the case of the gradient
percolation model, correlations occur over all length scales, and the
resulting distributions are fractal, at least up to the percolation
threshold.

In examining the changing development of Berlin in Fig. \ref{dy}a, it
appears that the fractal distribution remains quite stable over a period
of 70 years and this implies that any controls on growth that there
might have been do not show up in terms of the changing settlement
pattern, implying that the growth dynamics of the city are not
influenced by such control. A rather different test of such policies is
provided in the case of London where a green belt policy was first
established in the 1930s and rigorously enforced since the 1950s. The
question is whether this has been effective in changing the form of the
settlement pattern. First, it is not clear that the siting of new towns
beyond London's commuting field was ever beyond the percolation field
and thus it is entirely possible that the planned new settlements in the
1950s and 1960s based on existing village and town cores simply
reinforced the existing fractal pattern.

In the same manner, the imposition of local controls on growth in terms
of preserving green field land from development seems to have been based
on reinforcing the kind of spatial disorder consistent with morphologies
generated through correlated percolation. The regional green belt policy
was based on policies being defined locally and then aggregated into the
green belt itself, and this seems to suggest that the morphology of
nondevelopment that resulted was fractal. This is borne out in a fractal
analysis of development in the London region which suggests that the
policy has little impact on the overall morphology of the area
\cite{fractal,shepherd}. 
Moreover, we note that the coincidence between the settlement
area distribution for different cities and different years (Berlin 1920
and 1945, and London 1981) suggests that local planning policies such as
the green belt 
may have a relatively low impact on the distribution of
towns.  Our model suggests that the area distribution is determined by
the degree of interactions among development units, and that its scaling
properties are independent of time.  
Current debates on urban growth have now
shifted to the development of brownfield sites in cities, and it would
be interesting to quantify the extent to which such future developments
might reinforce or counter the ``natural'' growth of the city as implied
in these kinds of models.

To develop more detailed and conclusive insights into the impact of
urban policies on growth, it is necessary to develop the model
further. This model implies that the area and size distributions, the
degree of interaction amongst dependent units of development, and
fractal dimension are independent of time. The only time dependent
parameter is the gradient $\lambda$ and it appears that we might predict
future urban forms simply by extrapolating the value of $\lambda$ in
time. However, we have yet to investigate the influence of topography
and other physical constraints on development, the influence of
transport routes and the presence of several ``independent'' central
cores or CBDs in the urban region. 

These models can also be further adapted to predict bond as well as site
percolation and in future work we will explore the extent to which such
interactions between sites and cities might be modeled explicitly. Our
interest in such examples is in the universality of the exponents that
we have demonstrated here, and which we wish to relate to the impact of
urban planning policies.

\bigskip

\noindent{\small ACKNOWLEDGEMENTS:} We thank NSF and CNPq for
financial support.





\end{multicols}

\end{document}